# Quantum tricriticality at the superfluid-insulator transition of binary Bose mixtures

Yasuyuki Kato[1], Daisuke Yamamoto[2], and Ippei Danshita[3,4]
[1] *RIKEN Center for Emergent Matter Science (CEMS), Wako, Saitama 351-0198, Japan*
[2] *Condensed Matter Theory Laboratory, RIKEN, Wako, Saitama 351-0198, Japan*
[3] *Yukawa Institute for Theoretical Physics, Kyoto University, Kyoto 606-8502, Japan*
[4] *Computational Condensed Matter Physics Laboratory, RIKEN, Wako, Saitama 351-0198, Japan*

Quantum criticality near a tricritical point (TCP) is studied in the two-component Bose-Hubbard model on square lattices. The existence of quantum TCP on a boundary of superfluid-insulator transition is confirmed by quantum Monte Carlo simulations. Moreover, we analytically derive the quantum tricritical behaviors on the basis of an effective field theory. We find two significant features of the quantum tricriticality, that are its characteristic chemical potential dependence of the superfluid transition temperature and a strong density fluctuation. We suggest that these features are directly observable in existing experimental setups of Bose-Bose mixtures in optical lattices.

Rapid development in experiments with ultracold gases confined in optical lattices has advanced the studies of quantum phase transitions (QPTs), thanks to their precise controllability of various parameters, such as external potentials, interparticle interactions, and lattice geometry, over a wide range. Several QPTs that are of close relevance to other condensed matter systems have been realized in experiments, such as superfluid (SF)-Mott insulator (MI) transitions in a variety of lattice geometry [1–4], SF-Bose glass transitions in a random [5, 6] or quasi-periodic [7, 8] potential, magnetic transitions in a tilted [9] or triangular [10, 11] optical lattice, and topological transitions in a double-well optical lattice [12]. Recent experiments have reported even the observation of quantum critical behaviors accompanying the second-order QPT between vacuum and SF [13], thus providing new opportunities for studying quantum criticality in optical-lattice systems.

Tricriticality, or more generally, multicriticality is a fundamental concept in the study of phase transitions [14]. A tricritical point (TCP) marks a point at which a second-order (continuous) phase transition changes to a first-order (discontinuous) phase transition on a single phase boundary in a two parameter phase diagram. Tricriticality has been discussed in the contexts of several condensed matter systems, e.g., $FeCl_2$ [15], $^3$He-$^4$He mixture [16], and correlated electron matterials [17, 18], as well as in quantum chromodynamics [19]. Due to its unique nature unconventional critical properties are expected to – and indeed found to – emerge in the vicinity of a TCP. As such, exploration of TCPs can be a useful strategy for finding novel universality classes of phase transitions. Despite such ubiquity and importance of TCP, understanding of quantum tricriticality remains limited to a phenomenological level because of lack of experiments with flexible controllability and exact numerical simulations on a microscopic model, in contrast to classical one.

In this Letter, we use the unbiased numerical method of quantum Monte-Carlo (QMC) based on the Feynmann path integral [20] to show the existence of quantum TCPs in the ground state phase diagram of the two-component Bose-Hubbard model (BHM) on square lattices. This result suggests that quantum tricriticality can realistically be studied in Bose-Bose mixtures trapped in optical lattices, which are subsistent experimental setups [21–26]. From a simple mean-field (MF) analysis, we explain that effective two-body attraction causes the emergence of the TCPs. Furthermore, analyzing an effective continuum model, we derive critical behaviors regarding the TCP. In the finite-temperature phase diagrams obtained by the QMC method, we identify the quantum tricritical behavior of the SF-normal transition temperature, which can be directly measured in experiments [13].

We consider the two-component BHM on square lattices ($d = 2$) [27],

$$\mathcal{H} = -t \sum_{\alpha,\langle j,l\rangle} \left( b^\dagger_{\alpha,j} b_{\alpha,l} + \text{H.c.} \right) - \mu \sum_{\alpha,j} b^\dagger_{\alpha,j} b_{\alpha,j} + \sum_{\alpha,\alpha',j} \frac{U_{\alpha\alpha'}}{2} b^\dagger_{\alpha,j} b^\dagger_{\alpha',j} b_{\alpha',j} b_{\alpha,j}, \tag{1}$$

where $b^\dagger_{\alpha,j}$ ($b_{\alpha,j}$) is a creator (annihilator) of $\alpha$-type boson at site $j$, $t$ the nearest neighbor hopping amplitude, $\mu$ the chemical potential, and $U_{AA} = U_{BB} \equiv U > 0$ ($U_{AB} = U_{BA}$) the on-site intra-component (inter-component) repulsive interaction. This model describes Bose-Bose mixtures confined in an optical lattice [21–26]. In the optical-lattice experiments, $U/t$ is controlled by changing the depth of optical lattices [27], and $U_{AB}$ is controlled by using Feshbach resonances [23, 28, 29] or component-dependent optical lattices [26, 30]. Although the hopping amplitudes are different for each component in general, we use a common value $t$ for simplicity. Indeed, in a gas of $^{87}$Rb having two atomic spin states as an internal degree of freedom, the difference of hopping amplitudes between these two spin states is negligible [22].

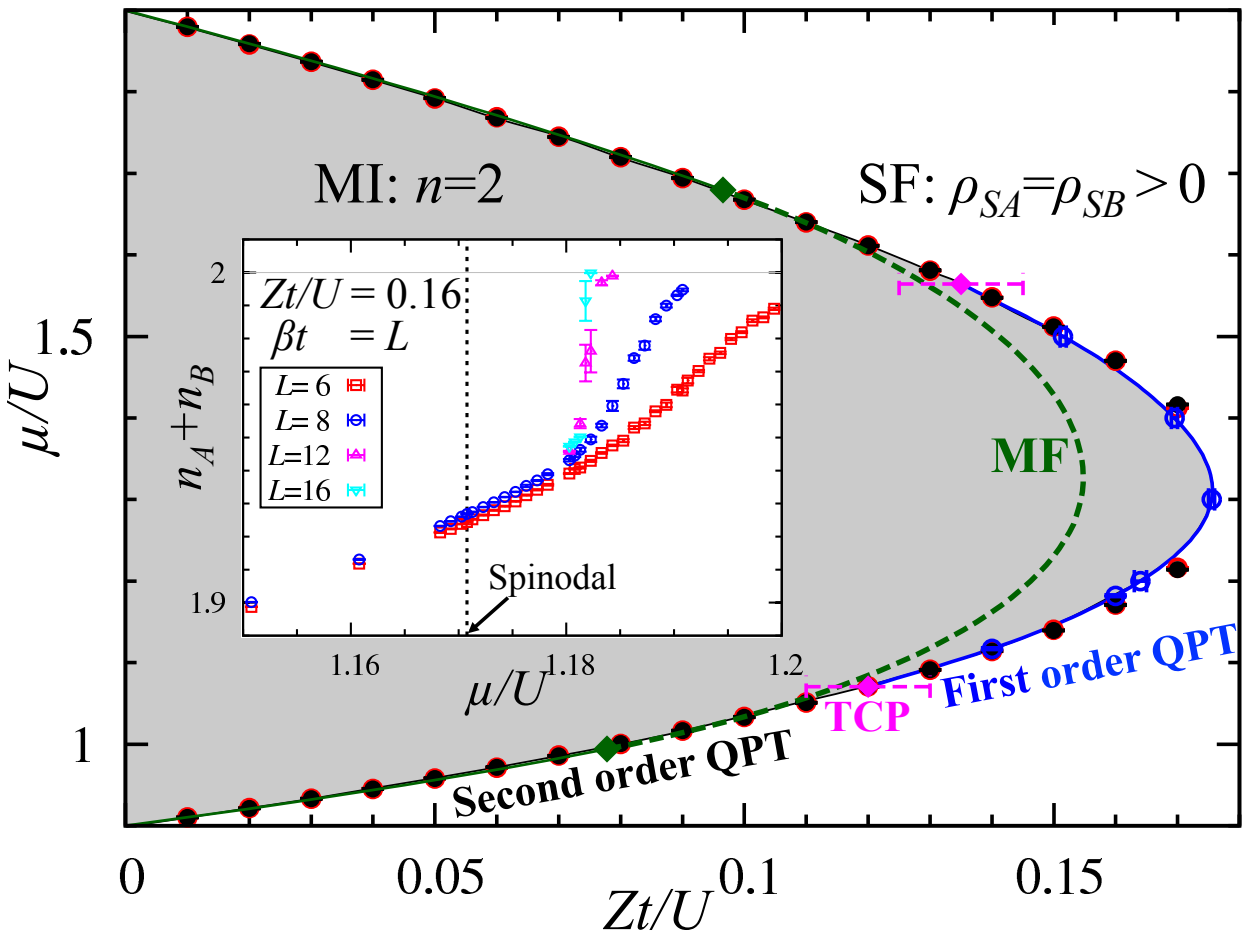

FIG. 1: (Color online) Ground state phase diagram at $U_{AB}/U = 0.9$ computed with QMC and MF. Thick (thin) green solid line shows phase boundary of first (second) order QPT computed by MF. Black (red) closed circles are the second order QPT points or spinodals computed by QMC simulations at $L = 16$ ($L = 12$) and $\mu/U = 1.35$. Blue open circles are the first order QPT points at $L = 16$. Diamonds are TCPs. Inset: $\mu$ dependence of total density of boson at $Zt/U = 0.16$.

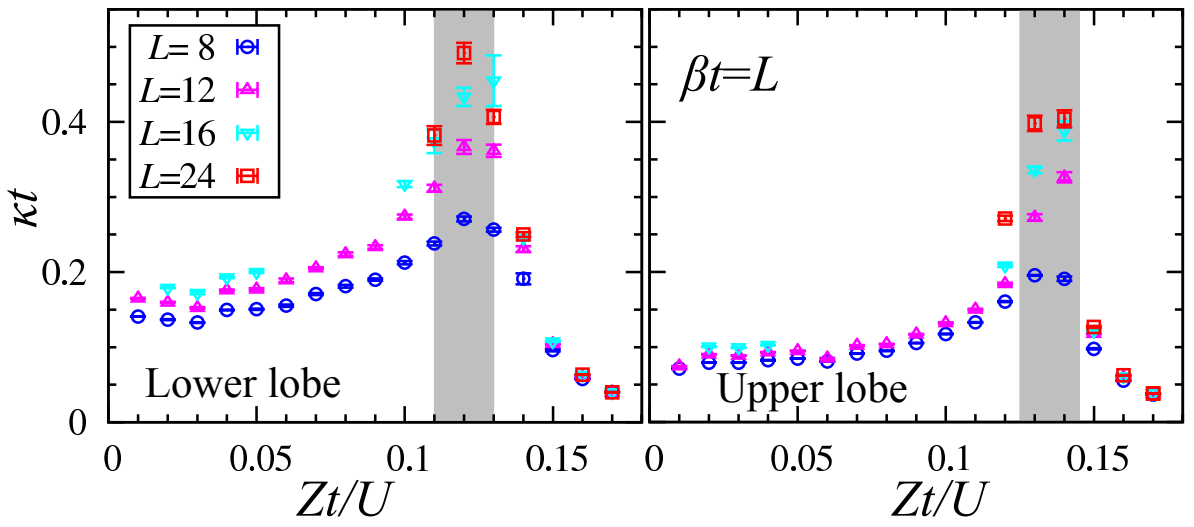

FIG. 2: (Color online) Density fluctuation $\kappa \equiv \partial n/\partial\mu$ at the lower (left panel) and upper (right panel) edges of the lobe estimated from the particle and hole gaps. The shaded regions mark the peaks of $\kappa$.

Notice that we choose $\hbar = k_{\rm B} = a = 1$ as our units throughout the paper, where $a$ is the lattice constant.

When $d \geq 2$, the possibility of the first-order QPT from SF to MI with even fillings has been previously suggested by Monte Carlo simulations on the two-component $J$-current model [31], which is a $(d+1)$-dimensional classical analog of Eq. (1), and MF analyses on Eq. (1) [32, 33]. Similar first-order QPTs have been found also in other related models [34, 35]. Below, we demonstrate the presence of the first-order QPT and the associated TCPs in the $t-\mu$ phase diagram by means of direct QMC simulations on Eq. (1).

We apply the worldline QMC method to the model (1) with a periodic boundary condition. We use a modified version [36] of the directed-loop algorithm [37] for updating worldline configurations. Although the modification is originally made for the single component BHM, it is also crucial for this application. We set the maximum occupation number at a single site as $n_{\rm max} = 4$ for each component.

Figure 1 shows the ground-state phase diagram of the model (1) near the $n = 2$ Mott lobe at $U_{AB}/U = 0.9$, where $n \equiv n_A + n_B$ is the total density and $n_\alpha$ is the density of type-$\alpha$ bosons. At this value of $U_{AB}/U$, the phase separation does not occur in either SF or MI phase. The phase boundary of second-order QPT between SF and MI is determined from the single-particle and single-hole Mott gaps estimated by using the QMC data of the Green's function $\langle b_{A,\boldsymbol{k}}(\tau) b^\dagger_{A,\boldsymbol{k}}(0)\rangle$ at $\boldsymbol{k} = 0$ [38], where $b^\dagger_{A,\boldsymbol{k}} \equiv L^{-d/2}\sum_j b^\dagger_{A,j}\exp(i\boldsymbol{k}\cdot\boldsymbol{x}_j)$ and $\boldsymbol{x}_j$ represents the position vector of site $j$. We estimate the Mott gaps at $\mu/U = 1.35$ and $L = 12$ and 16 with several values of $Zt/U$. In the scale of the phase diagram (Fig. 1), finite-size effects are negligible. In the case of the first-order QPT, the Mott gaps do not locate the phase boundary but the spinodal of the MI state. To identify the first-order phase boundary, we calculate the total density $n$ near the tip of the $n = 2$ Mott lobe as a function of $\mu/U$ or $Zt/U$ as shown in the inset of Fig. 1. At $Zt/U = 0.16$, we find a clear jump in $n$ versus $\mu/U$, and we determine the transition point from the position of the jump for $L = 16$. The spinodal located by the Mott gap is well separated from the true transition point and is located in the SF side. When $Zt/U$ decreases, the jump becomes smaller and is supposed to vanish at the TCP. However, it is practically very difficult to estimate numerically the position of TCPs from the vanishment of the jump, because the jump is too small to be detected close to TCPs.

Instead, we determine the TCPs from the difference in the critical behavior of the density fluctuation, $\kappa \equiv \partial n/\partial\mu$, between the generic [49] and tricritical transitions. Since the density of atoms is locally observable in optical-lattice experiments [39], this density fluctuation can be directly probed. To determine the critical behavior across the transition between MI with even filling and SF with incommensurate filling, we analyze the effective action in continuum given by

$$S^{\rm eff} = \beta V f_0 + \int d\tau \int d^dx \Bigg[ \sum_\alpha \left( \psi^*_\alpha \frac{\partial \psi_\alpha}{\partial\tau} + \frac{1}{2m}|\nabla\psi_\alpha|^2 - r_\alpha|\psi_\alpha|^2 + \frac{u}{2}|\psi_\alpha|^4 + \frac{w}{3}|\psi_\alpha|^6 \right) + u_{AB}|\psi_A|^2|\psi_B|^2 + w_{AB}(|\psi_A|^4|\psi_B|^2 + |\psi_A|^2|\psi_B|^4) \Bigg], \quad (2)$$

where $\beta \equiv T^{-1}$ is the inverse temperature, $V \equiv L^d$ the volume, and $f_0$ the free energy density of the MI state. A derivation of the effective model is shown in detail in the Supplemental Material [40]. In Eq. (2), $\psi_\alpha(\boldsymbol{x},\tau)$ denotes the SF order-parameter field of type-$\alpha$ bosons. Inclusion of terms up to the sixth order of $\psi_\alpha$ is necessary to describe the shift from generic transition to first-order one through the TCP. The dynamical and critical exponents for the transitions described by the action (2) are $z = 2$ and $\nu = 1/2$. Applying MF theory combined with renormalization group analysis to the effective model (2), we obtain the critical behaviors of the density versus the chemical potential as shown in Table I. The results imply that in two dimension $\kappa \propto \ln(1/\delta\mu)$ for the generic transition while $\kappa \propto \delta\mu^{-1/2}$ for the tricritical one (see Ref. [40] for a detailed derivation). Here $\delta\mu \equiv |\mu - \mu_c|$ and $\mu_c$ denotes $\mu$ at the critical point. It is remarkable that for the tricritical case there is no logarithmic correction even in $d = 2$ that is the upper critical dimension. Utilizing these critical behaviors, the TCPs are determined as follows. In Fig. 2, we depict $\kappa$ along the lower and upper edges of the lobe estimated by the Mott gap at $\mu/U = 1.35$ and $L = 16$. There we see that $\kappa$ has a distinct peak at a certain value of $Zt/U$. The peak position identifies the TCPs at which $\kappa$ diverges more strongly than at the generic QCPs or at the spinodal of the MI state. Notice that although $\kappa$ diverges weakly as $\sim \ln(1/\delta\mu)$ even at generic QCPs in $d = 2$, the QMC data do not show such a divergence because the phase boundary computed with $L = 16$ is expected to be located inside the SF phase in the thermodynamic limit.

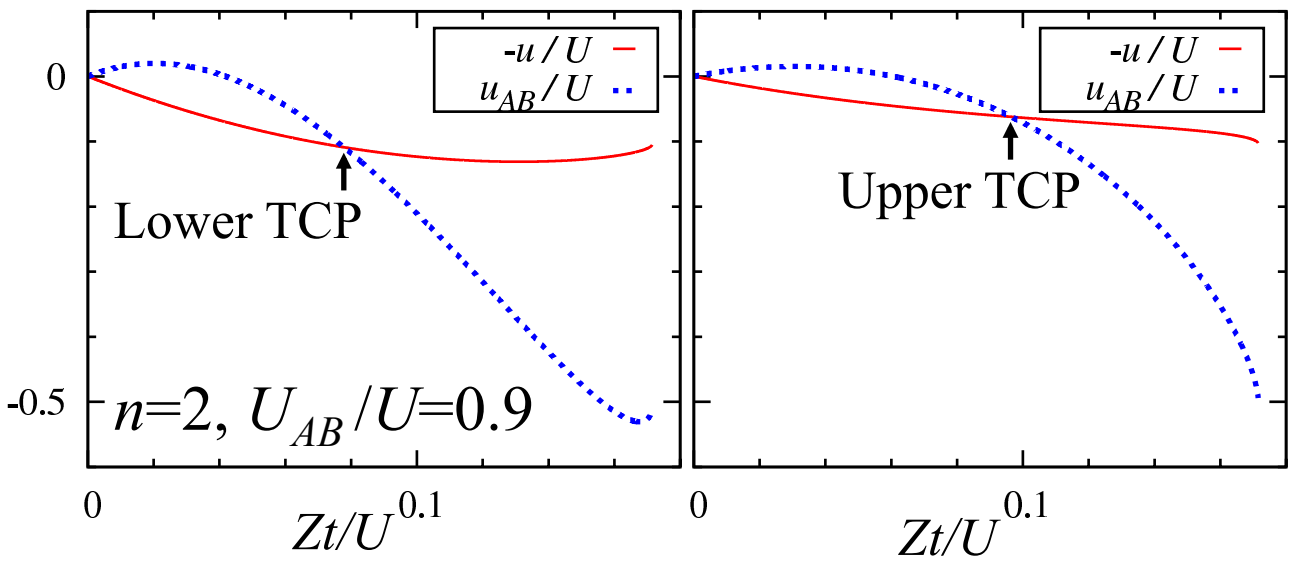

FIG. 3: (Color online) $Zt/U$ dependence of $-u/U$ and $u_{AB}/U$ at $U_{AB}/U = 0.9$, for the lower (left panel) and upper (right panel) edges of the lobe by Ginzburg-Landau expansion of energy density from $|n_A, n_B\rangle = |1, 1\rangle$.

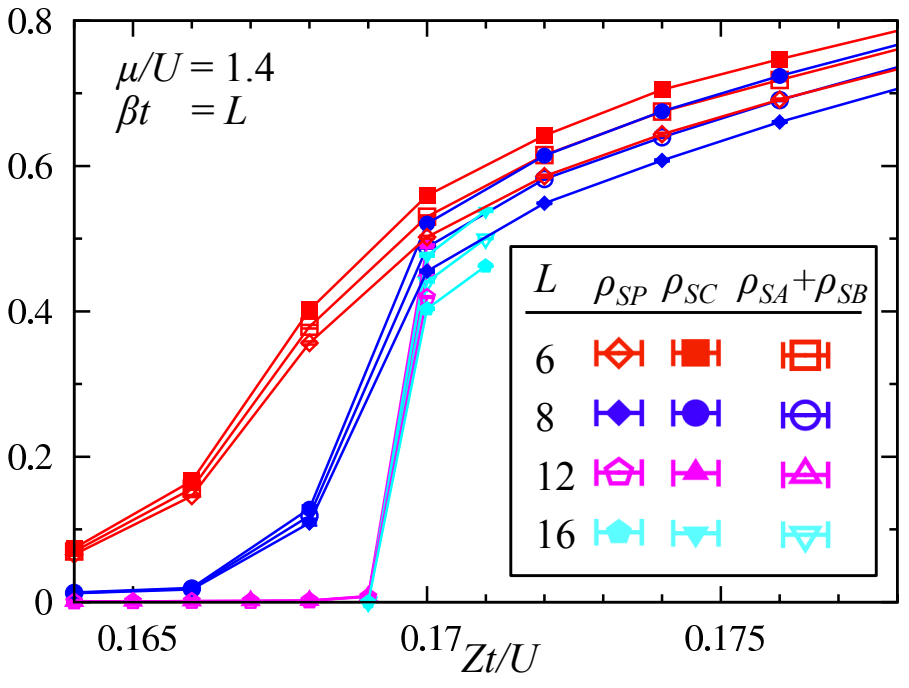

FIG. 4: (Color online) The SF stiffness $\rho_{SA} + \rho_{SB}$, $\rho_{SC}$, and $\rho_{SP}$ computed by QMC simulations at $\beta t = L$ and $\mu/U = 1.4$.

We discuss a physical mechanism for the shift of the QPT from second order to first order and the associated emergence of the TCP within a MF approximation. By the green line in Fig. 1, we show the phase boundary computed with the use of the MF theory [32, 33] for comparison. While quantitative difference in the first-order phase boundary and the TCP between MF and QMC is discernible, the MF analysis correctly captures the qualitative features of the phase diagram. Applying the MF approximation $\psi_\alpha(\boldsymbol{x}, \tau) = \phi_\alpha$ to the effective model (2), we obtain the MF action $S^{\rm mf} = \beta V f$ with the free energy density written as

$$f = f_0 - 2r\phi^2 + (u + u_{AB})\phi^4 + \frac{2}{3}(w + 3w_{AB})\phi^6, \quad (3)$$

where $r_A = r_B \equiv r$. Taking into account the symmetry between the two components, we here set $|\phi_A| = |\phi_B| \equiv \phi$ that corresponds to the superfluid order parameter $\langle b_{\alpha,j}\rangle$ characterizing the SF-MI transition. From Eq. (3), one sees that the transition is of first order when $u + u_{AB} < 0$ while it is of second order when $u + u_{AB} \geq 0$. In Fig. 3, we plot $-u$ and $u_{AB}$ as functions of $Zt/U$ along the lower and upper edges of the lobe (see Ref. [40] for how to calculate $u$ and $u_{AB}$). When $Zt/U$ increases in the region $Zt/U > 0.05$, $u_{AB}$ becomes strongly attractive so that $u + u_{AB}$ changes its sign at a certain point, which is nothing but the TCP. Thus, the shift of the QPT from second order to first order can be attributed to the strong inter-component attraction, which leads to the collapse of the SF state at low SF density.

To complete the ground-state phase diagram, one needs to reveal whether there exists the super-counterflow (SCF) order in the MI phase [31]. In Fig. 4, we show the SF stiffness computed with QMC from the fluctuation of winding numbers [41, 42] as $\rho_{S\alpha} \equiv \langle \boldsymbol{W}_\alpha^2\rangle/(4\beta t)$, $\rho_{SC} \equiv \langle(\boldsymbol{W}_A - \boldsymbol{W}_B)^2\rangle/(4\beta t)$, and $\rho_{SP} \equiv \langle(\boldsymbol{W}_A + \boldsymbol{W}_B)^2\rangle/(4\beta t)$, where $W_\alpha^{x(y)}$ is a winding number of $\alpha$-type bosons' worldlines of $x(y)$ direction. The SCF (paired superfluid) order is present if $\rho_{SC} > 0$ and $\rho_{SP} = 0$ ($\rho_{SP} > 0$ and $\rho_{SC} = 0$). All the kinds of stiffness vanish inside the MI region, thus confirming the absence of those SF orders. This validates the use of the effective action (2) that consists only of the order parameter of single-particle SF $\psi_\alpha$. In Fig. 4, we also find $\rho_{SC} > \rho_{SP}$, i.e., $\langle \boldsymbol{W}_A \cdot \boldsymbol{W}_B\rangle < 0$, that is consistent with the prediction of the effective attractive interaction $u_{AB} < 0$.

Next let us consider the finite-temperature phase diagram. While the nature of finite temperature transition differs from that at zero temperature, the transition temperature $T_c$ near a second-order QPT is governed by the quantum criticality [43, 44]. The critical behavior of $T_c$ is of particular importance in the sense that it has been directly measured in recent experiments in both contexts of $^4$He films [45] and optical lattices loaded with ultracold gases [13]. On the basis of the effective model (2), we derive the critical behavior of $T_c$ that is summarized in Table I (see Ref. [40] for details). As expected, the critical behaviors for the generic QPT is the same as that for the vacuum-SF QPT of a dilute Bose gas [44, 46, 47], and these QPTs belong to the same universality class. In contrast, the tricritical behavior is distinctly different from the standard one; specifically, $T_\mathrm{c} \propto \delta\mu^{1/d}$ if we disregard the logarithmic contributions in $d = 2$.

| | Density | Transition temperature |
|---|---|---|
| $d=2$ Generic | $\delta\mu \ln(1/\delta\mu)$ | $\delta\mu \ln(1/\delta\mu) \,/\, \ln\ln(1/\delta\mu)$ |
| $d=2$ Tricritical | $\delta\mu^{1/2}$ | $\delta\mu^{1/2} / \ln(1/\delta\mu)$ |
| $d>2$ Generic | $\delta\mu$ | $\delta\mu^{2/d}$ |
| $d>2$ Tricritical | $\delta\mu^{1/2}$ | $\delta\mu^{1/d}$ |

TABLE I: Quantum criticality of the density $\delta n = |n - n_0|$ measured on that at the MI state $n_0$ and the transition temperature $T_\mathrm{c}$ as functions of the chemical potential measured on the critical point $\delta\mu \equiv |\mu - \mu_\mathrm{c}|$.

Figure 5 shows the transition temperatures as functions of $\mu/U$ obtained from QMC simulations for several values of $Zt/U$. The transition temperatures are estimated from jumps of $\rho_{SA}$ and $n$. If a transition to SF is of the Berezinskii-Kosterlitz-Thouless (BKT) type, it is well-known that $\rho_{SA}$ exhibits the universal jump, $\Delta\rho_{SA} = T/(t\pi)$, at the transition point [48]. Thus the BKT transition temperature is estimated from the cross-point of $T/(t\pi)$ and $\rho_{SA}(L = 48)$ as shown in Fig. 5(b). On the other hand, $\rho_{SA}$ shows a larger jump than $T/(t\pi)$ at the first-order phase transition. As shown in Fig. 5(c), we find the first-order phase transition at finite temperatures. This means that there is a tricritical line in the $T - t - \mu$ phase diagram and the quantum TCPs are its end points.

To examine the shift of the criticality from $T_c \propto \delta\mu$ (generic) to $\propto \delta\mu^{1/2}$ (tricritical), we fit $T_c/t$ by assuming a fitting function $T_\mathrm{fit}(\delta\bar{\mu}) = [C_1^2 + C_2\delta\bar{\mu}]^{1/2} - C_1$ with fitting parameters $C_{1,2}$, where $\delta\bar{\mu} \equiv \delta\mu/U$. Since $T_\mathrm{fit} \propto \delta\bar{\mu}^{1/2}$ for $C_1 = 0$ while $T_\mathrm{fit} \propto \delta\bar{\mu}$ for $C_1 > 0$ at small $\delta\bar{\mu}$, the approach to the tricritical behavior can be identified as the decrease of $C_1$ towards zero. By the fitting to the numerical data in Fig. 5(a), we indeed find that $C_1$ at the TCP, i.e. $Zt/U = 0.12$, is much closer to zero compared to $C_1$ for the generic QPTs ($Zt/U = 0.04$ and 0.08). Thus, the tricritical behavior of $T_c$ has been corroborated. Notice that within the temperature range of our numerical simulations the fitting function that neglects logarithmic contributions fits better to the data than the one with logarithmic contributions, as was also the case in previous experiments [13, 45].

In conclusion, we have computed the ground-state and finite-temperature phase diagrams near the $n = 2$ Mott lobe of the two-component BHM in square lattices by the QMC method. It was shown that the SF-MI transition is of first order near the tip of the Mott lobe while it is of second order far from the tip. We have identified the TCPs on the SF-MI phase boundary. Since the model is a quantitative counterpart of a realistic experimental system, namely a Bose-Bose mixture in an optical lattice, this finding of the TCPs makes it possible for optical-lattice experiments to address the issue of quantum tricriticality. Moreover, we have derived critical behaviors of the QPT across the TCP. The predicted quantum tricritical behavior of the SF-normal transition temperature may be examined in future experiments.

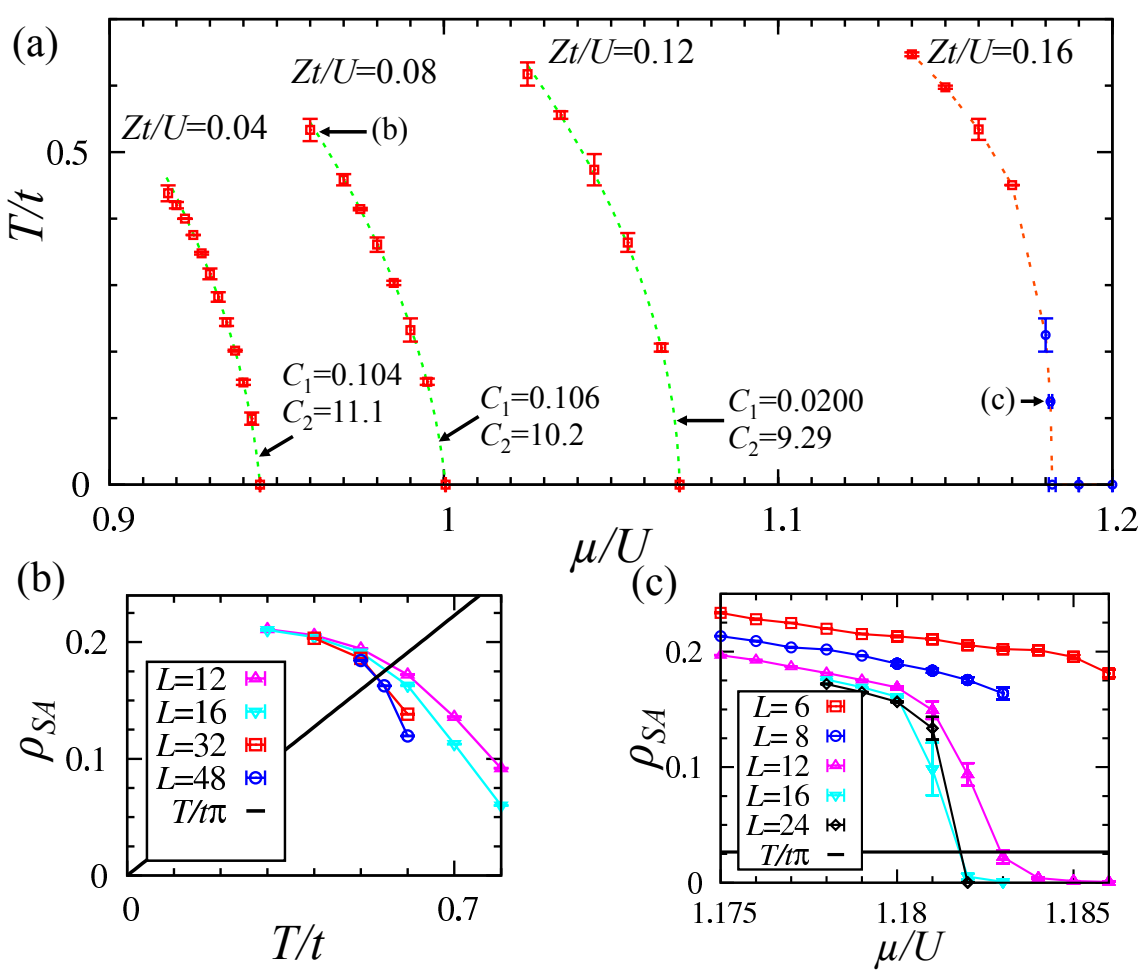

FIG. 5: (Color online) Results of QMC simulations at finite temperature. (a) Transition temperatures at $Zt/U = 0.04, 0.08, 0.12$ and 0.16. Circles (blue) and squares (red) represent the first-order and BKT transitions, respectively. (b) Temperature dependence of $\rho_{SA}$ at $Zt/U = 0.08$ and $\mu/U = 0.96$. (c) Chemical potential dependence of $\rho_{SA}$ at $Zt/U = 0.16$ and $\beta t = 12$.

We thank P. Sengupta, G. Marmorini, T. Sato, and K. Totsuka for useful comments and discussions. Numerical calculations were conducted on the RIKEN Integrated Cluster of Clusters (RICC). I. D. acknowledges support from KAKENHI Grants No. 25800228 and No. 25220711.

# Supplementary Material for "Quantum Tricriticality at the Superfluid-Insulator Transition of Binary Bose Mixtures"

Yasuyuki Kato[1], Daisuke Yamamoto[2], and Ippei Danshita[3,4]
[1]RIKEN Center for Emergent Matter Science (CEMS), Wako, Saitama 351-0198, Japan
[2]Condensed Matter Theory Laboratory, RIKEN, Wako, Saitama 351-0198, Japan
[3]Yukawa Institute for Theoretical Physics, Kyoto University, Kyoto 606-8502, Japan
[4]Computational Condensed Matter Physics Laboratory, RIKEN, Wako, Saitama 351-0198, Japan

In this supplementary material, we present a detailed derivation of the effective continuum model that is Eq. (2) of the main text. We also explain in detail how to derive the critical scaling of the density and the transition temperature for the superfluid-insulator transition, which are summarized in Table I of the main text, from the effective model.

## I. EFFECTIVE CONTINUUM MODEL

We consider the system of the two-component Bose-Hubbard model that is given in Eq. (1) of the main text. In the path integral representation, the partition function of this model is given by

$$\Xi = \mathrm{Tr}\, e^{-\beta \hat{H}} = \int \mathcal{D}b_A^* \mathcal{D}b_A \mathcal{D}b_B^* \mathcal{D}b_B \exp(-S[b_A, b_A^*, b_B, b_B^*]/\hbar). \tag{1}$$

Here the Euclidean action is written as

$$S[b_A, b_A^*, b_B, b_B^*] = S_A + S_B + S_{AB}, \tag{2}$$

where

$$S_\alpha = \int_{-\frac{\hbar\beta}{2}}^{\frac{\hbar\beta}{2}} d\tau \left[ \sum_j b_{\alpha,j}^* \left( \hbar \frac{\partial}{\partial \tau} - \mu_\alpha + \frac{U_{\alpha\alpha}}{2} b_{\alpha,j}^* b_{\alpha,j} \right) b_{\alpha,j} - \sum_{j,l} t_{\alpha,jl}\, b_{\alpha,j}^* b_{\alpha,l} \right], \tag{3}$$

$$S_{AB} = \int_{-\frac{\hbar\beta}{2}}^{\frac{\hbar\beta}{2}} d\tau \sum_j U_{AB}\, b_{A,j}^* b_{A,j} b_{B,j}^* b_{B,j}. \tag{4}$$

We assume that the hopping process is allowed only between nearest neighboring sites such that the element of the hopping matrix $\hat{t}_\alpha$ takes the form that

$$t_{\alpha,jl} = \begin{cases} t_\alpha, & \text{if } j \text{ and } l \text{ are nearest neighboring} \\ 0, & \text{otherwise} \end{cases}. \tag{5}$$

While the Bose-Hubbard parameters in the main text are taken to be symmetric with respect to the replacement $A \leftrightarrow B$ (e.g. $t_A = t_B \equiv t$ and $U_{AA} = U_{BB} \equiv U$), we here do not assume the symmetry in order to be more general until we derive the effective action. We are especially interested in the phase transition between the Mott insulator (MI) with even filling and the superfluid (SF) with incommensurate filling upon varying the chemical potential $\mu_\alpha$. When the system is in a close vicinity of the transition, the correlation length is much larger than the underlying lattice spacing $a$ so that one can describe the system with use of the effective continuum model [1]. In this section, we derive the effective action, following the prescription used in Ref. [2] for the case of the single-component Bose-Hubbard model.

Inserting the auxiliary fields $\Psi_\alpha$ and $\Psi_\alpha^*$ via the Stratnovich-Hubbard transformation, the action becomes

$$S[b_A, b_A^*, b_B, b_B^*, \Psi_A, \Psi_A^*, \Psi_B, \Psi_B^*] = \int d\tau \sum_{\alpha,j,l} (\hat{t}_\alpha^{-1})_{jl} \Psi_{\alpha,j}^* \Psi_{\alpha,l} + S^{(0)} + S^{\mathrm{pert}} \tag{6}$$

where

$$S^{(0)} = \int d\tau \sum_{\alpha,j} b_{\alpha,j}^* \left( \hbar \frac{\partial}{\partial \tau} - \mu_\alpha + \frac{U_{\alpha\alpha}}{2} b_{\alpha,j}^* b_{\alpha,j} \right) b_{\alpha,j} + S_{AB}, \tag{7}$$

$$S^{\mathrm{pert}} = -\int d\tau \sum_{\alpha,j} (b_{\alpha,j}^* \Psi_{\alpha,j} + \Psi_{\alpha,j}^* b_{\alpha,j}), \tag{8}$$

We next integrate out the fields $b_{\alpha,j}$ and $b^*_{\alpha,j}$, and the action is formally expressed as

$$S^{\rm eff}[\Psi_A,\Psi_A^*,\Psi_B,\Psi_B^*]=\hbar\beta V f_0+\int d\tau\sum_{\alpha,j,l}(\hat{t}_\alpha^{-1})_{jl}\Psi^*_{\alpha,j}\Psi_{\alpha,l}-\hbar\ln\Big\langle\exp\left(-S^{\rm pert}/\hbar\right)\Big\rangle_0, \quad (9)$$

where

$$\langle O\rangle_0\equiv\int\mathcal{D}b_A^*\mathcal{D}b_A\mathcal{D}b_B^*\mathcal{D}b_B\; O\exp\left(-S^{(0)}/\hbar\right), \quad (10)$$

and $V\equiv(La)^d$ is the volume of the system. $S^{(0)}$ describes the system of completely decoupled sites, which is already diagonalized such that the eigenenergies and the eigenstates can be expressed in a simple analytical form. Hence, one can compute the average $\langle O\rangle_0$ with the operator $O$ consisting of a product of $b_{\alpha,j}$ and $b^*_{\alpha,j}$. In order to construct the effective action more explicitly, we first perform a cumulant expansion up to the sixth order with respect to the fields $\Psi_{\alpha,j}(\tau)$ and $\Psi^*_{\alpha,j}(\tau)$. We next take the continuum limit $aj\to\boldsymbol{x}$ and $a^{-d/2}\Psi_{\alpha,j}(\tau)\to\Psi_\alpha(\boldsymbol{x},\tau)$, and finally express the order parameter in the dimension of the wave function as $\psi_\alpha\equiv\Psi_\alpha/(Zt_\alpha)$, where $Z=2d$ is the coordination number. In this way, we obtain the effective action,

$$S^{\rm eff}[\psi_A,\psi_A^*,\psi_B,\psi_B^*]=\beta V f_0+S_A^{\rm eff}+S_B^{\rm eff}+S_{AB}^{\rm eff}, \quad (11)$$

where

$$S_\alpha^{\rm eff} = \int d\tau\int d^dx\left[\hbar K_\alpha\psi_\alpha^*\frac{\partial\psi_\alpha}{\partial\tau}+\hbar^2J_\alpha\left|\frac{\partial\psi_\alpha}{\partial\tau}\right|^2+\frac{\hbar^2}{2m_\alpha}|\nabla\psi_\alpha|^2-r_\alpha|\psi_\alpha|^2+\frac{u_\alpha}{2}|\psi_\alpha|^4+\frac{w_\alpha}{3}|\psi_\alpha|^6\right], \quad (12)$$

$$S_{AB}^{\rm eff} = \int d\tau\int d^dx\left[u_{AB}|\psi_A|^2|\psi_B|^2+w_{AB}|\psi_A|^4|\psi_B|^2+w_{BA}|\psi_A|^2|\psi_B|^4\right]. \quad (13)$$

The coefficients $K_\alpha$, $J_\alpha$, $m_\alpha$, $r_\alpha$, $u_\alpha$, $u_{AB}$, $w_\alpha$, $w_{AB}$, and $w_{BA}$ are functions of the parameters in the original Hubbard-type model, namely the hopping energy $t_\alpha$, the chemical potential $\mu_\alpha$, and interparticle interactions $U_{\alpha\alpha'}$. Here, we assume that the ground state of the decoupled system $S^{(0)}$ is given by a product of local Fock states $|\nu_A,\nu_B\rangle=|g,g\rangle$ in order to discuss the transition between the MI with even filling $2g$ and the SF. From the second-order cumulant expansion, we obtain the explicit form of $K_\alpha$, $J_\alpha$, $m_\alpha$, and $r_\alpha$ as

$$K_\alpha = (Zt_\alpha)^2\left(\frac{g+1}{(E_\alpha^{(+)}-E_{g,g})^2}-\frac{g}{(E_\alpha^{(-)}-E_{g,g})^2}\right), \quad (14)$$

$$J_\alpha = (Zt_\alpha)^2\left(\frac{g+1}{(E_\alpha^{(+)}-E_{g,g})^3}+\frac{g}{(E_\alpha^{(-)}-E_{g,g})^3}\right), \quad (15)$$

$$m_\alpha = \frac{\hbar^2}{2t_\alpha a^2}, \quad (16)$$

$$r_\alpha = -Zt_\alpha+(Zt_\alpha)^2\left(\frac{g+1}{E_\alpha^{(+)}-E_{g,g}}+\frac{g}{E_\alpha^{(-)}-E_{g,g}}\right), \quad (17)$$

where

$$E_{\nu_A,\nu_B} \equiv \sum_\alpha\left(-\mu_\alpha\nu_\alpha+\frac{U_{\alpha\alpha}}{2}\nu_\alpha(\nu_\alpha-1)\right)+U_{AB}\nu_A\nu_B, \quad (18)$$

$$E_\alpha^{(\pm)} \equiv \begin{cases} E_{g\pm1,g}, & \text{if } \alpha=A\\ E_{g,g\pm1}, & \text{if } \alpha=B\end{cases}. \quad (19)$$

While the coefficients of the higher order terms, namely $u_\alpha$, $u_{AB}$, $w_\alpha$, $w_{AB}$, and $w_{BA}$, can be obtained also via the cumulant expansion of Eq. (9), it is more straightforward to use alternatively the perturbative mean-field expansion [3, 4].

From Eqs. (6-8), the corresponding effective Hamiltonian is written as

$$\mathcal{H}^{\rm eff}=\sum_{\alpha,j,l}(\hat{t}_\alpha^{-1})_{jl}\Psi^*_{\alpha,j}\Psi_{\alpha,l}+\sum_j\mathcal{H}_j^{(0)}+\sum_{\alpha,j}V_{\alpha,j} \quad (20)$$

where

$$\mathcal{H}_j^{(0)} = -\mu \sum_{\alpha} b_{\alpha,j}^{\dagger} b_{\alpha,j} + \sum_{\alpha,\alpha'} \frac{U_{\alpha\alpha'}}{2} b_{\alpha,j}^{\dagger} b_{\alpha',j}^{\dagger} b_{\alpha',j} b_{\alpha,j}, \tag{21}$$

$$V_{\alpha,j} = -(b_{\alpha,j}^{\dagger}\Psi_{\alpha,j} + \Psi_{\alpha,j}^{*} b_{\alpha,j}). \tag{22}$$

The coefficients of the effective action in Eqs. (11-13) are obtained by the standard perturbation method in which the last term $V_{\alpha,j}$ coupled with the fields $\Psi_{\alpha,j}$ and $\Psi_{\alpha,j}^{*}$ is considered as a small perturbation on the unperturbed Hamiltonian $\mathcal{H}_j^{(0)}$. The coefficients $u_\alpha$ and $u_{AB}$ are derived from the fourth-order perturbation energy as

$$\begin{aligned}
&\sum_{\nu_A,\nu_B}{}' \sum_{\lambda_A,\lambda_B}{}' \frac{|\langle g,g|V_j|\nu_A,\nu_B\rangle|^2 |\langle g,g|V_j|\lambda_A,\lambda_B\rangle|^2}{(E_{\nu_A,\nu_B}-E_{g,g})(E_{\lambda_A,\lambda_B}-E_{g,g})^2} \\
&-\sum_{\nu_A,\nu_B}{}' \sum_{\lambda_A,\lambda_B}{}' \sum_{\rho_A,\rho_B}{}' \frac{\langle g,g|V_j|\nu_A,\nu_B\rangle\langle \nu_A,\nu_B|V_j|\lambda_A,\lambda_B\rangle\langle \lambda_A,\lambda_B|V_j|\rho_A,\rho_B\rangle\langle \rho_A,\rho_B|V_j|g,g\rangle}{(E_{\nu_A,\nu_B}-E_{g,g})(E_{\lambda_A,\lambda_B}-E_{g,g})(E_{\rho_A,\rho_B}-E_{g,g})} \\
&= \left(\hat{V}_j^{(1)\dagger}\hat{G}^{(1)}\hat{V}_j^{(1)}\right)\left(\hat{V}_j^{(1)\dagger}(\hat{G}^{(1)})^2\hat{V}_j^{(1)}\right) - \hat{V}_j^{(1)\dagger}\hat{G}^{(1)}\hat{V}_j^{(2)\dagger}\hat{G}^{(2)}\hat{V}_j^{(2)}\hat{G}^{(1)}\hat{V}_j^{(1)} \\
&= \sum_{\alpha}\frac{u_\alpha a^{-d}}{2}\left|\frac{\Psi_{\alpha,j}}{Zt_\alpha}\right|^4 + u_{AB}a^{-d}\left|\frac{\Psi_{A,j}}{Zt_A}\right|^2\left|\frac{\Psi_{B,j}}{Zt_B}\right|^2,
\end{aligned} \tag{23}$$

where $V_j = \sum_\alpha V_{\alpha,j}$ and the sum $\sum'_{\nu_A,\nu_B}$ runs over all the eigenstates of $\mathcal{H}_j^0$ other than the initial state $|\nu_A,\nu_B\rangle = |g,g\rangle$. The matrices in the third line of Eq. (23) are defined by

$$\begin{aligned}
\hat{V}_j^{(1)} &= \begin{pmatrix} \sqrt{g}\Psi_{A,j}^{*} & \sqrt{g+1}\Psi_{A,j} & \sqrt{g}\Psi_{B,j}^{*} & \sqrt{g+1}\Psi_{B,j} \end{pmatrix}^T, \\
\hat{G}^{(1)} &= \left(\mathrm{diag}(E_{g-1,g}, E_{g+1,g}, E_{g,g-1}, E_{g,g+1}) - E_{g,g}\hat{I}_4\right)^{-1}, \\
\hat{V}_j^{(2)} &= \begin{pmatrix}
\sqrt{g-1}\Psi_{A,j}^{*} & \sqrt{g}\Psi_{B,j}^{*} & 0 & \sqrt{g}\Psi_{B,j} & 0 & 0 & 0 & 0 \\
0 & 0 & 0 & 0 & \sqrt{g}\Psi_{B,j}^{*} & \sqrt{g+2}\Psi_{A,j} & \sqrt{g+1}\Psi_{B,j} & 0 \\
0 & \sqrt{g}\Psi_{A,j}^{*} & \sqrt{g-1}\Psi_{B,j}^{*} & 0 & \sqrt{g+1}\Psi_{A,j} & 0 & 0 & 0 \\
0 & 0 & 0 & \sqrt{g}\Psi_{A,j}^{*} & 0 & 0 & \sqrt{g+1}\Psi_{A,j} & \sqrt{g+2}\Psi_{B,j}
\end{pmatrix}^T, \\
\hat{G}^{(2)} &= \left(\mathrm{diag}(E_{g-2,g}, E_{g-1,g-1}, E_{g,g-2}, E_{g-1,g+1}, E_{g+1,g-1}, E_{g+2,g}, E_{g+1,g+1}, E_{g,g+2}) - E_{g,g}\hat{I}_8\right)^{-1},
\end{aligned}$$

where $\hat{I}_n$ denotes the $n\times n$ identity matrix. From Eq. (23), the explicit forms of $u_\alpha$ and $u_{AB}$ are given by

$$\begin{aligned}
u_\alpha &= 2a^d(Zt_\alpha)^4\left[\left(\frac{g+1}{E_\alpha^{(+)}-E_{g,g}} + \frac{g}{E_\alpha^{(-)}-E_{g,g}}\right)\left(\frac{g+1}{(E_\alpha^{(+)}-E_{g,g})^2} + \frac{g}{(E_\alpha^{(-)}-E_{g,g})^2}\right)\right. \\
&\quad \left. -\left(\frac{(g+1)(g+2)}{(E_\alpha^{(+)}-E_{g,g})^2(E_\alpha^{(2+)}-E_{g,g})} + \frac{g(g+1)}{(E_\alpha^{(-)}-E_{g,g})^2(E_\alpha^{(2-)}-E_{g,g})}\right)\right]
\end{aligned} \tag{24}$$

$$\begin{aligned}
u_{AB} &= a^dZ^4t_A^2t_B^2\left[\left(\frac{g+1}{E_A^{(+)}-E_{g,g}} + \frac{g}{E_A^{(-)}-E_{g,g}}\right)\left(\frac{g+1}{(E_B^{(+)}-E_{g,g})^2} + \frac{g}{(E_B^{(-)}-E_{g,g})^2}\right)\right. \\
&\quad +\left(\frac{g+1}{E_B^{(+)}-E_{g,g}} + \frac{g}{E_B^{(-)}-E_{g,g}}\right)\left(\frac{g+1}{(E_A^{(+)}-E_{g,g})^2} + \frac{g}{(E_A^{(-)}-E_{g,g})^2}\right) \\
&\quad -\left(\frac{1}{E_A^{(+)}-E_{g,g}} + \frac{1}{E_B^{(+)}-E_{g,g}}\right)^2 \frac{(g+1)^2}{E_{AB}^{(++)}-E_{g,g}} \\
&\quad -\left(\frac{1}{E_A^{(+)}-E_{g,g}} + \frac{1}{E_B^{(-)}-E_{g,g}}\right)^2 \frac{g(g+1)}{E_{AB}^{(+-)}-E_{g,g}} \\
&\quad -\left(\frac{1}{E_A^{(-)}-E_{g,g}} + \frac{1}{E_B^{(+)}-E_{g,g}}\right)^2 \frac{g(g+1)}{E_{AB}^{(-+)}-E_{g,g}}
\end{aligned}$$

$$
-\left(\frac{1}{E_A^{(-)}-E_{g,g}}+\frac{1}{E_B^{(-)}-E_{g,g}}\right)^2\frac{g^2}{E_{AB}^{(--)}-E_{g,g}}\Bigg], \tag{25}
$$

where

$$
E_\alpha^{(2\pm)} \equiv \begin{cases} E_{g\pm2,g}, & \text{if } \alpha=A \\ E_{g,g\pm2}, & \text{if } \alpha=B \end{cases}, \tag{26}
$$

$$
E_{AB}^{(\pm\pm)} \equiv E_{g\pm1,g\pm1}, \tag{27}
$$

$$
E_{AB}^{(\pm\mp)} \equiv E_{g\pm1,g\mp1}. \tag{28}
$$

In a similar fashion, the coefficients $w_\alpha$, $w_{AB}$, and $w_{BA}$ can be obtained from the sixth-order perturbation:

$$
\begin{aligned}
&\left(\hat{V}_j^{(1)\dagger}\hat{G}^{(1)}\hat{V}_j^{(2)\dagger}\hat{G}^{(2)}\hat{V}_j^{(2)}\hat{G}^{(1)}\hat{V}_j^{(1)}\right)\left(\hat{V}_j^{(1)\dagger}(\hat{G}^{(1)})^2\hat{V}_j^{(1)}\right)-\left(\hat{V}_j^{(1)\dagger}\hat{G}^{(1)}\hat{V}_j^{(1)}\right)\left(\hat{V}_j^{(1)\dagger}(\hat{G}^{(1)})^2\hat{V}_j^{(1)}\right)^2\\
&-\left(\hat{V}_j^{(1)\dagger}\hat{G}^{(1)}\hat{V}_j^{(1)}\right)\left(\hat{V}_j^{(1)\dagger}(\hat{G}^{(1)})^2\hat{V}_j^{(2)\dagger}\hat{G}^{(2)}\hat{V}_j^{(2)}\hat{G}^{(1)}\hat{V}_j^{(1)}+\hat{V}_j^{(1)\dagger}\hat{G}^{(1)}\hat{V}_j^{(2)\dagger}(\hat{G}^{(2)})^2\hat{V}_j^{(2)}\hat{G}^{(1)}\hat{V}_j^{(1)}\right.\\
&\left.+\hat{V}_j^{(1)\dagger}\hat{G}^{(1)}\hat{V}_j^{(2)\dagger}\hat{G}^{(2)}\hat{V}_j^{(2)}(\hat{G}^{(1)})^2\hat{V}_j^{(1)}\right)-\left(\hat{V}_j^{(1)\dagger}\hat{G}^{(1)}\hat{V}_j^{(1)}\right)^2\left(\hat{V}_j^{(1)\dagger}(\hat{G}^{(1)})^3\hat{V}_j^{(1)}\right)\\
&-\hat{V}_j^{(1)\dagger}\hat{G}^{(1)}\hat{V}_j^{(2)\dagger}\hat{G}^{(2)}\hat{V}_j^{(3)\dagger}\hat{G}^{(3)}\hat{V}_j^{(3)}\hat{G}^{(2)}\hat{V}_j^{(2)}\hat{G}^{(1)}\hat{V}_j^{(1)}\\
&=\sum_\alpha\frac{w_\alpha a^{-2d}}{3}\left|\frac{\Psi_{\alpha,j}}{Zt_\alpha}\right|^6+w_{AB}a^{-2d}\left|\frac{\Psi_{A,j}}{Zt_A}\right|^4\left|\frac{\Psi_{B,j}}{Zt_B}\right|^2+w_{BA}a^{-2d}\left|\frac{\Psi_{A,j}}{Zt_A}\right|^2\left|\frac{\Psi_{B,j}}{Zt_B}\right|^4,
\end{aligned} \tag{29}
$$

where

$$
\hat{V}_j^{(3)}=\begin{pmatrix}
\sqrt{g-2}\Psi_{A,j}^* & 0 & 0 & 0 & 0 & 0 & 0 & 0\\
\sqrt{g}\Psi_{B,j}^* & \sqrt{g-1}\Psi_{A,j}^* & 0 & 0 & 0 & 0 & 0 & 0\\
0 & \sqrt{g-1}\Psi_{B,j}^* & \sqrt{g}\Psi_{A,j}^* & 0 & 0 & 0 & 0 & 0\\
0 & 0 & \sqrt{g-2}\Psi_{B,j}^* & 0 & 0 & 0 & 0 & 0\\
\sqrt{g+1}\Psi_{B,j} & 0 & 0 & \sqrt{g-1}\Psi_{A,j}^* & 0 & 0 & 0 & 0\\
\sqrt{g-1}\Psi_{A,j} & \sqrt{g}\Psi_{A,j} & 0 & \sqrt{g+1}\Psi_{B,j}^* & 0 & 0 & 0 & 0\\
0 & \sqrt{g}\Psi_{B,j} & \sqrt{g-1}\Psi_{B,j} & 0 & \sqrt{g+1}\Psi_{A,j}^* & 0 & 0 & 0\\
0 & 0 & \sqrt{g+1}\Psi_{A,j} & 0 & \sqrt{g-1}\Psi_{B,j}^* & 0 & 0 & 0\\
0 & 0 & 0 & 0 & \sqrt{g+2}\Psi_{A,j} & \sqrt{g}\Psi_{B,j}^* & 0 & 0\\
0 & 0 & 0 & 0 & \sqrt{g}\Psi_{B,j} & \sqrt{g+2}\Psi_{A,j}^* & \sqrt{g+1}\Psi_{B,j}^* & 0\\
0 & 0 & 0 & \sqrt{g}\Psi_{A,j} & 0 & 0 & \sqrt{g+1}\Psi_{A,j}^* & \sqrt{g+2}\Psi_{B,j}^*\\
0 & 0 & 0 & \sqrt{g+2}\Psi_{B,j} & 0 & 0 & 0 & \sqrt{g}\Psi_{A,j}^*\\
0 & 0 & 0 & 0 & 0 & \sqrt{g+3}\Psi_{A,j} & 0 & 0\\
0 & 0 & 0 & 0 & 0 & \sqrt{g+1}\Psi_{B,j} & \sqrt{g+2}\Psi_{A,j} & 0\\
0 & 0 & 0 & 0 & 0 & 0 & \sqrt{g+2}\Psi_{B,j} & \sqrt{g+1}\Psi_{A,j}\\
0 & 0 & 0 & 0 & 0 & 0 & 0 & \sqrt{g+3}\Psi_{B,j}
\end{pmatrix},
$$

$$
\begin{aligned}
\hat{G}^{(3)}=\Big(&\mathrm{diag}(E_{g-3,g},E_{g-2,g-1},E_{g-1,g-2},E_{g,g-3},E_{g-2,g+1},E_{g-1,g},E_{g,g-1},E_{g+1,g-2},E_{g+2,g-1},E_{g+1,g},E_{g,g+1},\\
&E_{g-1,g+2},E_{g+3,g},E_{g+2,g+1},E_{g+1,g+2},E_{g,g+3})-E_{g,g}\hat{I}_{16}\Big)^{-1}.
\end{aligned}
$$

We do not present the final analytical formula for $w_\alpha$, $w_{AB}$, and $w_{BA}$ herein since they are rather lengthy. Note that, for $g\le2$, the terms including $E_{\nu_A,\nu_B}$ with $\nu_A<0$ or $\nu_B<0$ should be omitted in Eqs. (23-29). We use Eqs. (24) and (25) to plot $u$ and $u_{AB}$ along the SF-MI phase boundaries at $g=1$ in Fig. 3 of the main text.

The transition natures described by the action (11) significantly depend on whether the coefficient of the first-order time derivative is $K_\alpha=0$ or $K_\alpha\neq0$ [1]. When $K_\alpha=0$, the density does not change in the transition process, meaning that the transition occurs between the MI and the SF with commensurate filling. In this case, the $J_\alpha$ term is the leading dynamical contribution to the action such that the dynamical exponent of the transition is $z=1$. On the other hand, the case $K_\alpha\neq0$ corresponds to the transition between the MI and the SF with incommensurate filling, which is accompanied by a density change, and its dynamical exponent is $z=2$. As was mentioned above, we are especially interested in the latter case, and we will consider only the latter case hereafter.

We can simplify further the effective action (11) as follows. Since the $J_\alpha$ term is irrelevant with respect to the $K_\alpha$ term, the $J_\alpha$ term can be dropped in the action. Moreover, we take the unit in which $K_\alpha = 1$. Then, $S_\alpha^{\rm eff}$ is rewritten as

$$S_\alpha^{\rm eff} = \int d\tau \int d^d x \left[ \hbar \psi_\alpha^* \frac{\partial \psi_\alpha}{\partial \tau} + \frac{\hbar^2}{2m_\alpha} |\nabla \psi_\alpha|^2 - r_\alpha |\psi_\alpha|^2 + \frac{u_\alpha}{2} |\psi_\alpha|^4 + \frac{w_\alpha}{3} |\psi_\alpha|^6 \right]. \tag{30}$$

Thus, the action (11) with Eqs. (13) and (30) is the effective continuum model that describes the transition between the MI with even filling and the SF with incommensurate filling in the system of the two-component Bose-Hubbard model. Notice that this effective action can not be used to describe the transition between SF and MI with odd filling, in which the super-counterflow order is also involved [5]. In Fig. 4 of the main text, it is shown that such super-counterflow order is absent in the case of even filling.

Since we assume that the Hubbard parameters are symmetric with respect to the replacement $A \leftrightarrow B$ as $t_A = t_B (\equiv t)$, $\mu_A = \mu_B (\equiv \mu)$, $U_{AA} = U_{BB} (\equiv U)$, and $U_{AB} = U_{BA}$, the coefficients in the effective model possess the same symmetry as $m_A = m_B$, $r_A = r_B$, $u_A = u_B (\equiv u)$, $w_A = w_B (\equiv w)$, and $w_{AB} = w_{BA}$. However, for evaluating the transition temperature one needs to calculate the polarizability $P \equiv \frac{\partial n_-}{\partial \mu_-}$ that requires the expression of the free energy density with $r_A \neq r_B$, where $n_\pm \equiv n_A \pm n_B$, $\mu_\pm \equiv \mu_A \pm \mu_B$, and $n_\alpha$ denotes the density of the $\alpha$-type particles. Hence, we here write $r_A$ and $r_B$ independently under the assumption that $r_- \ll r_+$, where $r_\pm = r_A \pm r_B$. While inclusion of the terms up to the fourth order of $\psi_\alpha$ is sufficient to describe the critical behaviors of the generic transition [6], the sixth order terms are necessary to capture the change to the first-order transition and the associated tricriticality [4]. The quantity $f_0$ is the free energy density of a system with zero hopping energy, in which lattice sites are completely decoupled from one another. The derivative of $f_0$ with respect to $\mu_\alpha$ gives the filling factor $\nu_\alpha$ of the Mott insulator phase as

$$\frac{\partial f_0}{\partial \mu_\alpha} = -\frac{\nu_\alpha}{a^d}. \tag{31}$$

In order to simplify the notation, we take the lattice spacing $a$ and the energy $\epsilon_a \equiv \frac{\hbar^2}{ma^2}$ as units of the length and the energy, and express the action in a dimensionless form,

$$\tilde{S}^{\rm eff}[\psi_A, \psi_A^*, \psi_B, \psi_B^*] = \tilde{\beta} \tilde{V} \tilde{f}_0 + \tilde{S}_A^{\rm eff} + \tilde{S}_B^{\rm eff} + \tilde{S}_{AB}^{\rm eff}, \tag{32}$$

where

$$\tilde{S}_\alpha^{\rm eff} = \int_{-\frac{\tilde{\beta}}{2}}^{\frac{\tilde{\beta}}{2}} d\tilde{\tau} \int d^d \tilde{x} \left[ \tilde{\psi}_\alpha^* \frac{\partial \tilde{\psi}_\alpha}{\partial \tilde{\tau}} + \frac{1}{2} |\tilde{\nabla} \tilde{\psi}_\alpha|^2 - \tilde{r}_\alpha |\tilde{\psi}_\alpha|^2 + \frac{\tilde{u}}{2} |\tilde{\psi}_\alpha|^4 + \frac{\tilde{w}}{3} |\tilde{\psi}_\alpha|^6 \right], \tag{33}$$

$$\tilde{S}_{AB}^{\rm eff} = \int_{-\frac{\tilde{\beta}}{2}}^{\frac{\tilde{\beta}}{2}} d\tilde{\tau} \int d^d \tilde{x} \left[ \tilde{u}_{AB} |\tilde{\psi}_A|^2 |\tilde{\psi}_B|^2 + \tilde{w}_{AB} (|\tilde{\psi}_A|^4 |\tilde{\psi}_B|^2 + |\tilde{\psi}_A|^2 |\tilde{\psi}_B|^4) \right]. \tag{34}$$

Specifically, the dimensionless quantities are related to their original values through

$$\begin{aligned} \tilde{S}^{\rm eff} &= S^{\rm eff}/\hbar,\ \tilde{\beta} = \beta \times \epsilon_a,\ \tilde{V} = V/a^d,\ \tilde{f}_0 = f_0 \times a^d/\epsilon_a,\ \tilde{\tau} = \tau \times \epsilon_a/\hbar, \\ \tilde{\boldsymbol{x}} &= \boldsymbol{x}/a,\ \tilde{\psi} = \psi \times a^{d/2},\ \tilde{r}_\alpha = r_\alpha/\epsilon_a,\ \tilde{u} = u/(\epsilon_a a^d), \tilde{w} = w/(\epsilon_a a^{2d}). \end{aligned} \tag{35}$$

Henceforth, we omit the tildes written above the quantities for simplicity.

## II. MEAN-FIELD APPROXIMATION

In the next section, we will see that $d = 2$ is the upper critical dimension regardless of whether the transition is generic or tricritical. This means that the critical behaviors at $d > 2$ obeys scaling formulae derived with a mean-field approximation while there are some logarithmic corrections to the mean-field scaling at $d = 2$. In either case the mean-field scaling formulae are necessary, and we derive them in this section.

Applying a mean field approximation, $\psi_\alpha(\boldsymbol{x}, \tau) = \phi_\alpha$, the action is written as

$$S_{\rm mf} = \beta V f. \tag{36}$$

with the free energy density,

$$f = f_0 - r_A \phi_A^2 - r_B \phi_B^2 + \frac{u}{2}(\phi_A^4 + \phi_B^4) + u_{AB} \phi_A^2 \phi_B^2 + \frac{w}{3}(\phi_A^6 + \phi_B^6) + w_{AB}(\phi_A^4 \phi_B^2 + \phi_A^2 \phi_B^4). \tag{37}$$

The conditions $\frac{\partial f}{\partial \phi_A}=0$, $\frac{\partial f}{\partial \phi_B}=0$ leads to the two solutions

$$(\phi_A^2,\phi_B^2) \;=\; (0,0), \tag{38}$$

$$(\phi_A^2,\phi_B^2) \;\simeq\; \left(\varphi+\frac{r_-}{2(u_-+w_-\varphi)},\varphi-\frac{r_-}{2(u_-+w_-\varphi)}\right), \tag{39}$$

either of which is a ground state. Here $\varphi$ corresponds to the order parameter value at $r_A=r_B$, which is given by

$$\varphi=\frac{\sqrt{u_+^2+2r_+w_+}-u_+}{2w_+}. \tag{40}$$

In Eqs. (39) and (40), we introduced the notations $u_\pm=u\pm u_{AB}$, $w_+=w+3w_{AB}$, and $w_-=w-w_{AB}$. The former solution describes the MI phase and the latter corresponds to the SF phase. Notice that the latter solution is approximately obtained under the assumption that $r_-\ll r_+$.

Substituting Eq. (39) into Eq. (37), we obtain the free energy density for the superfluid phase,

$$f_{\rm sf}\simeq f_0+\frac{u_+^3}{6w_+^2}+\frac{r_+u_+}{2w_+}-\frac{1}{6w_+^2}(u_+^2+2r_+w_+)^{3/2}-\frac{r_-^2}{4(u_-+w_-\varphi)}. \tag{41}$$

From the free energy density Eq. (41), one can see that when $u_+\geq 0$, the second order (continuous) transition from MI to SF occurs at $r_+=0$ upon increasing $r_+$ [4]. When $u_+<0$, the transition is first order (discontinuous) and occurs at $r_+=-3u_+^2/(8w_+)$. This means that the transition at $u_+=0$ corresponds to the tricritical point.

We first consider the case of the generic transition, i.e. $u_+>0$. Assuming $u_+^2\gg r_+w_+$, the free energy density is expressed as

$$f_{\rm sf}\simeq f_0-\frac{r_+^2}{4u_+}-\frac{r_-^2}{4u_-}. \tag{42}$$

Here we are interested in the behavior of $f_{\rm sf}$ in the vicinity of the critical point $(\mu_A,\mu_B)=(\mu_{\rm c},\mu_{\rm c})$, and we expand the parameters to the leading order around the critical point,

$$r_\alpha \;\simeq\; \sum_\gamma D_{\alpha\gamma}\delta\mu_\gamma, \tag{43}$$

$$u_\pm \;\simeq\; u_{\pm,{\rm c}}, \tag{44}$$

$$w_\pm \;\simeq\; w_{\pm,{\rm c}}, \tag{45}$$

where

$$D_{\alpha\gamma}\equiv\left.\frac{\partial r_\alpha}{\partial\mu_\gamma}\right|_{(\mu_A,\mu_B)=(\mu_{\rm c},\mu_{\rm c})}, \tag{46}$$

$$u_{\pm,{\rm c}}\equiv u_\pm|_{(\mu_A,\mu_B)=(\mu_{\rm c},\mu_{\rm c})}, \tag{47}$$

$$w_{\pm,{\rm c}}\equiv w_\pm|_{(\mu_A,\mu_B)=(\mu_{\rm c},\mu_{\rm c})}. \tag{48}$$

(49)

Since $D_{AA}=D_{BB}(\equiv D)$ and $D_{AB}=D_{BA}$, the free energy density is rewritten as

$$f_{\rm sf}\simeq f_0-\frac{D_+^2}{4u_{+,{\rm c}}}\delta\mu_+^2-\frac{D_-^2}{4u_{-,{\rm c}}}\delta\mu_-^2. \tag{50}$$

where $D_\pm=D\pm D_{AB}$. Using Eq. (50) and the thermodynamic relations, one can obtain the total density

$$n_+\equiv n_A+n_B=-\frac{\partial f_{\rm sf}}{\partial\mu_A}-\frac{\partial f_{\rm sf}}{\partial\mu_B}\simeq 2g+\frac{D_+^2}{u_{+,{\rm c}}}\delta\mu_+, \tag{51}$$

and the relative density

$$n_-\equiv n_A-n_B=-\frac{\partial f_{\rm sf}}{\partial\mu_A}+\frac{\partial f_{\rm sf}}{\partial\mu_B}\simeq\frac{D_-^2}{u_{-,{\rm c}}}\delta\mu_-. \tag{52}$$

Differentiating the densities with respect to the chemical potentials leads to the density fluctuation

$$\kappa = 2\frac{\partial n_+}{\partial \mu_+} \simeq \frac{2D_+^2}{u_{+,\mathrm{c}}}, \tag{53}$$

and the polarizability

$$P = \frac{\partial n_-}{\partial \mu_-} \simeq \frac{D_-^2}{u_{-,\mathrm{c}}}. \tag{54}$$

We next consider the case of the tricritical transition ($u_+ = 0$), in which the free energy density is given by

$$f_{\mathrm{sf}} \simeq -\frac{\sqrt{2}}{3}\frac{r_+^{3/2}}{w_+^{1/2}} - \frac{r_-^2}{4u_-}. \tag{55}$$

Substituting Eqs. (43), (44), and (45) into Eq. (55), one obtains

$$f_{\mathrm{sf}} \simeq f_0 - \frac{\sqrt{2}}{3}\frac{D_+^{3/2}}{w_{+,\mathrm{c}}^{1/2}}\delta\mu_+^{3/2} - \frac{D_-^2}{4u_{-,\mathrm{c}}}\delta\mu_-^2. \tag{56}$$

From this free energy density, one can easily calculate the total density

$$n_+ = -\frac{\partial f_{\mathrm{sf}}}{\partial \mu_A} - \frac{\partial f_{\mathrm{sf}}}{\partial \mu_B} \simeq 2g + \frac{\sqrt{2}D_+^{3/2}}{w_{+,\mathrm{c}}^{1/2}}\delta\mu_+^{1/2}, \tag{57}$$

and the density fluctuation

$$\kappa = 2\frac{\partial n_+}{\partial \mu_+} \simeq \frac{\sqrt{2}D_+^{3/2}}{w_{+,\mathrm{c}}^{1/2}}\delta\mu_+^{-1/2}. \tag{58}$$

Notice that the expressions of the relative density $n_-$ and the polarizability $P$ for the tricritical transition are the same as those for the generic transition, namely Eqs. (52) and (54).

## III. RENORMALIZATION GROUP AND LOGARITHMIC CORRECTIONS

In the previous section, we have obtained the scaling formulae of the critical behaviors within the mean-field approximation. However, they are valid only above the upper critical dimension $d_{\mathrm{c}}$, and at $d = d_{\mathrm{c}}$ one needs to take into account some logarithmic corrections to the scaling formulae. In this section, we present a renormalization group (RG) analysis in order to show that $d_{\mathrm{c}} = 2$ and to derive the logarithmic corrections. We do not explicitly discuss the case of $d < 2$ because the first order SF-MI transition has never been reported in previous exact numerical analyses on the two-component Bose-Hubbard model at $d = 1$.

The renormalization group equations for the action Eq. (32) are given by

$$\frac{dr_\alpha}{dl} = 2r_\alpha, \tag{59}$$

$$\frac{du}{dl} = (2-d)u - K_d u^2, \tag{60}$$

$$\frac{du_{AB}}{dl} = (2-d)u_{AB} - K_d u_{AB}^2, \tag{61}$$

$$\frac{dw}{dl} = 2(1-d)w, \tag{62}$$

$$\frac{dw_{AB}}{dl} = 2(1-d)w_{AB}, \tag{63}$$

where $K_d = 2/[(4\pi)^{d/2}\Gamma(d/2)]$ and $\Gamma(x)$ is the Gamma function. Equations (59), (60), and (61) have been previously derived in Ref. [7]. From the RG equations, we see that the most relevant coefficient is $r_\alpha$ and its scaling dimension is 2. Hence, the critical exponent, which is defined as the inverse of the scaling dimension of the most relevant coefficient [1], is $\nu = 1/2$ for both generic and tricritical transitions. When $d > 2$, the interaction terms $u$, $u_{AB}$, $w$, and $w_{AB}$ are all irrelevant so that the mean-field scaling formulae are valid; thus, the upper critical dimension is $d_{\rm c} = 2$. Especially, we see from Eqs. (51) and (57) with $\delta\mu_- = 0$ that the formulae of the total density for the generic and tricritical transitions at $d > 2$ are given by $|\delta n_+| \propto |\delta\mu|$ and $|\delta n_+| \propto |\delta\mu|^{1/2}$, respectively, where $\delta n_+ \equiv n_+ - 2g$. These formulae are listed in Table I of the main text.

When $d = 2$, the terms $u$ and $u_{AB}$ are marginally irrelevant while $w$ and $w_{AB}$ are still irrelevant. In this case, the solution of the renormalization group equations can be analytically expressed as

$$r_\alpha(l) = e^{2l} r_\alpha(0), \tag{64}$$

$$u(l) = \frac{u(0)}{1 + \frac{u(0)}{4\pi} \ln \left| \frac{r_+(l)}{r_+(0)} \right|}, \tag{65}$$

$$u_{AB}(l) = \frac{u_{AB}(0)}{1 + \frac{u_{AB}(0)}{4\pi} \ln \left| \frac{r_+(l)}{r_+(0)} \right|}, \tag{66}$$

$$w(l) = e^{2(1-d)l} w(0), \tag{67}$$

$$w_{AB}(l) = e^{2(1-d)l} w_{AB}(0). \tag{68}$$

The renormalization group flow will be interrupted on the scale $l = l_*$ that satisfies $r_+(l_*) = 1$ [7]. At $l = l_*$, the system is out of the critical region, and the free energy density is safely given by its mean-field value of Eq. (41). When the system is sufficiently close to the transition point ($r_+(0) = 0$) so that $r_+(0) \ll e^{-4\pi/u(0)}$, $e^{-4\pi/u_{AB}(0)}$, the logarithmic term becomes dominant and one obtains

$$u_+(l_*) \simeq \frac{8\pi}{\ln \frac{1}{|r_+(0)|}}, \tag{69}$$

$$u_-(l_*) \simeq \left( \frac{4\pi}{\ln \frac{1}{|r_+(0)|}} \right)^2 \left( -\frac{1}{u(0)} + \frac{1}{u_{AB}(0)} \right). \tag{70}$$

The free energy density with the bare parameters is related to that with the renormalized parameters as [8]

$$f_{\rm sf}(l = 0) = e^{-(d+z)l_*} f_{\rm sf}(l = l_*), \tag{71}$$

where $z = 2$ is the dynamical exponent.

In the case of the generic transition ($u_+ > 0$), substituting Eqs. (64), (67), (68), (69), and (70) into Eq. (42), one obtains

$$f_{\rm sf}(l = 0) \simeq -\frac{1}{32\pi} D_+^2 \delta\mu_+^2 \ln \frac{1}{|\delta\mu_+|} - \frac{1}{(16\pi)^2} \frac{u_+^2 - u_-^2}{u_-} D_-^2 \delta\mu_-^2 \left( \ln \frac{1}{|\delta\mu_+|} \right)^2 \tag{72}$$

This free energy density allows us to calculate the total density

$$n_+ \simeq 2g + \frac{1}{8\pi} D_+^2 \delta\mu_+ \ln \frac{1}{|\delta\mu_+|} - \frac{1}{(8\pi)^2} \frac{u_+^2 - u_-^2}{u_-} D_-^2 \frac{\delta\mu_-^2}{\delta\mu_+} \ln \frac{1}{|\delta\mu_+|}, \tag{73}$$

the relative density

$$n_- \simeq \frac{1}{(8\pi)^2} \frac{u_+^2 - u_-^2}{u_-} D_-^2 \delta\mu_- \left( \ln \frac{1}{|\delta\mu_+|} \right)^2, \tag{74}$$

the density fluctuation

$$\kappa \simeq \frac{1}{4\pi} D_+^2 \ln \frac{1}{|\delta\mu_+|} + \frac{1}{32\pi^2} \frac{u_+^2 - u_-^2}{u_-} D_-^2 \frac{\delta\mu_-^2}{\delta\mu_+^2} \ln \frac{1}{|\delta\mu_+|}. \tag{75}$$

and the polarizability

$$P \simeq \frac{1}{(8\pi)^2} \frac{u_+^2 - u_-^2}{u_-} D_-^2 \left( \ln \frac{1}{|\delta\mu_+|} \right)^2. \tag{76}$$

In the case of the tricritical transition $u_+ = 0$, substituting Eqs. (64), (67), (68), (69), and (70) into Eq. (55), one obtains

$$f_{\rm sf}(l=0) \simeq -\frac{\sqrt{2}}{3}\frac{D_+^{3/2}}{w_{+,{\rm c}}^{1/2}}\delta\mu_+^{3/2} - \frac{1}{(16\pi)^2}\frac{u_+^2-u_-^2}{u_-}D_-^2\delta\mu_-^2\left(\ln\frac{1}{|\delta\mu_+|}\right)^2 \tag{77}$$

From this free energy density, one can calculate the total density

$$n_+ \simeq 2g + \sqrt{2}\frac{D_+^{3/2}}{w_{+,{\rm c}}^{1/2}}\delta\mu_+^{1/2} - \frac{1}{(8\pi)^2}\frac{u_+^2-u_-^2}{u_-}D_-^2\frac{\delta\mu_-^2}{\delta\mu_+}\ln\frac{1}{|\delta\mu_+|}, \tag{78}$$

and the density fluctuation

$$\kappa \simeq \sqrt{2}\frac{D_+^{3/2}}{w_{+,{\rm c}}^{1/2}}\delta\mu_+^{-1/2} + \frac{1}{32\pi^2}\frac{u_+^2-u_-^2}{u_-}D_-^2\frac{\delta\mu_-^2}{\delta\mu_+^2}\ln\frac{1}{|\delta\mu_+|}. \tag{79}$$

Notice that the expressions of the relative density $n_-$ and the polarizability $P$ for the tricritical transition are the same as those for the generic transition, namely Eqs. (74) and (76).

Setting $\delta\mu_- = 0$ in Eqs. (73) and (78), we see that the total densities scale with the chemical potential as $|\delta n_+| \propto |\delta\mu|\ln\frac{1}{|\delta\mu|}$ for the generic transition and $|\delta n_+| \propto |\delta\mu|^{1/2}$ for the tricritical transition, which are listed in Table I of the main text. The scaling formulae of $n_+$ and $\kappa$ for the generic transition of the two-component Bose-Hubbard model at $d \geq 2$ coincides with those for the single-component case [1], and thus the two transitions belong to the same universality class as expected. In the tricritical case, as seen in Eqs. (78) and (79), there is no logarithmic correction for $n_+$ and $\kappa$ at $\delta\mu_- = 0$ even in two dimension, and this property is in clear contrast to the generic transition.

## IV. TRANSITION TEMPERATURE

When the temperature increases in the superfluid phase, a transition to the normal fluid phase occurs at a certain temperature, which we refer to as the transition temperature $T_{\rm c}$. In this section, we derive the scaling formulae of the transition temperatures in the vicinity of the SF-MI transition at zero temperature for both generic and tricritical cases.

We first consider the case of $d > 2$. In this case, the Bose-Einstein condensation (BEC) can occur at finite temperatures, and the superfluid transition is accompanied by the BEC transition at the same temperature. In the vicinity of the SF-MI transition, the density of mobile particles (or holes) is so low that the transition temperature obeys the scaling formula for the ideal Bose gas [9],

$$T_{\rm c} \propto \delta n_+^{2/d}. \tag{80}$$

Substituting the scaling formulae of the density derived in the previous sections into Eq. (80), we express the transition temperature in terms of the chemical potential

$$T_{\rm c} \propto \begin{cases} |\delta\mu|^{2/d}, & \text{for the generic transition} \\ |\delta\mu|^{1/d}, & \text{for the tricritical transition} \end{cases}, \tag{81}$$

This result is shown in Table I of the main text.

We next consider the case of $d = 2$, in which BEC does not exist at finite temperatures in the thermodynamics limit, as known as the Hohenberg-Marmin-Wagner theorem [10, 11], and there is no relation between the BEC and superfluid transition temperatures unlike in $d > 2$. To evaluate $T_{\rm c}$ at $d = 2$, we follow the prescription shown in Ref. [6], where $T_{\rm c}$ is determined by the condition that the normal fluid density $n_{\rm nm}$ is equal to the total density $n_+$ at the transition temperature. In the two-component Bose gas, the normal fluid fraction is given by

$$n_{\rm nm} = \sum_{\lambda=+,-}\frac{1}{2}\int\frac{d^2p}{(2\pi)^2}p^2\left(-\frac{dN_{p,\lambda}}{d\epsilon_\lambda}\right) \tag{82}$$

where

$$N_{p,\lambda} = \frac{1}{e^{\beta\epsilon_\lambda(p)}-1} \tag{83}$$

is the Bose distribution function with the excitation spectra of the two-component Bose gas [12]

$$\epsilon_{+}(p) = \sqrt{\frac{p^2}{2}\left(\frac{p^2}{2}+\frac{n_+}{\kappa}\right)}, \tag{84}$$

$$\epsilon_{-}(p) = \sqrt{\frac{p^2}{2}\left(\frac{p^2}{2}+\frac{n_+}{P}\right)}. \tag{85}$$

Imposing the condition $n_{\rm nm} = n_+$ at $T = T_{\rm c}$ and assuming $|\delta\mu| \ll T_{\rm c}$ lead to [13]

$$T_{\rm c} = 2\pi n_+/(\ln\kappa + \ln P). \tag{86}$$

In the case of the generic transition, substituting Eqs. (73), (75), and (76) with $\delta\mu_- = 0$ into Eq. (86), we obtain the scaling formula of the transition temperature,

$$T_{\rm c} \propto |\delta\mu| \ln\frac{1}{|\delta\mu|} / \ln\ln\frac{1}{|\delta\mu|}. \tag{87}$$

In the tricritical case, substituting Eqs. (78), (79), and (76) with $\delta\mu_- = 0$ into Eq. (86), we obtain

$$T_{\rm c} \propto |\delta\mu|^{1/2} / \ln\frac{1}{|\delta\mu|}. \tag{88}$$

These scaling formulae of the transition temperature are listed in the Table I of the main text.